\documentclass[submission,copyright,creativecommons]{eptcs}
\providecommand{\event}{ACL2 2025} 

\usepackage{iftex}

\usepackage{bytefield}
\usepackage{listings}
\usepackage{xcolor}
\usepackage{xurl}
\usepackage{breakurl}
\usepackage[capitalise,noabbrev,nameinlink]{cleveref}
\usepackage{tikz}

\usetikzlibrary{arrows, positioning}
\tikzstyle{block} = [rectangle, draw, fill=blue!20, 
    text width=5em, text centered, rounded corners, minimum height=4em]
\tikzstyle{line} = [draw, -latex']

\definecolor{GrayCodeBlock}{RGB}{248,248,248}
\definecolor{BlackText}{RGB}{110,107,94}
\definecolor{RedTypename}{RGB}{182,86,17}
\definecolor{GreenString}{RGB}{96,172,57}
\definecolor{PurpleKeyword}{RGB}{184,84,212}
\definecolor{GrayComment}{RGB}{170,170,170}
\definecolor{GoldDocumentation}{RGB}{180,165,45}

 \lstdefinestyle{progstyle}{
     aboveskip=1pt,
     belowskip=1pt,
     showstringspaces=false,
     columns=flexible,
     basicstyle={\footnotesize\ttfamily},
     numbersep=1pt,
     numberstyle=\tiny\color{gray},
     keywordstyle=\color{blue},
     commentstyle=\color{green},
     stringstyle=\color{purple},
     breaklines=true,
     breakatwhitespace=true,
     tabsize=8,
     keepspaces=true
   }

\lstdefinelanguage{acl2}
{
    columns=fullflexible,
    keepspaces=true,
    frame=single,
    framesep=0pt,
    framerule=0pt,
    framexleftmargin=4pt,
    framexrightmargin=4pt,
    framextopmargin=5pt,
    framexbottommargin=3pt,
    xleftmargin=4pt,
    xrightmargin=4pt,
    backgroundcolor=\color{GrayCodeBlock},
    basicstyle=\scriptsize\ttfamily\color{BlackText},
    keywords={
        define,defun,defthm,defmacro,defconst,defstobj,defattach,b*,
        let,let*,lambda,if,cond,case,and,or,not,
        implies,
        declare,local,encapsulate,include-book,
        in-package,defpackage,
        prove,proof,qed,skip-proofs,
        theorem,lemma,corollary,
        def-gl-thm, def-fgl-thm,
        gl::def-gl-thm, fgl::def-fgl-thm
    },
    keywordstyle=\color{PurpleKeyword},
    ndkeywords={
        t,nil,state,
        part\-select,\:low,\:width,unless,
        verify-guards,
        equal,eql,
        car,cdr,cons,consp,
        symbolp,numberp,stringp,
        acl2-numberp,rationalp,integerp,
        member,assoc,
        list,append,reverse
        def-gl-thm, def-fgl-thm,
    },
    ndkeywordstyle=\color{RedTypename},
    comment=[l][\color{GrayComment}\slshape]{;},
    morecomment=[s][\color{GrayComment}\slshape]{\#|}{|\#},
    stringstyle=\color{GreenString},
    string=[b]"
}

\ifpdf
  \usepackage{underscore}         
  \usepackage[T1]{fontenc}        
\else
  \usepackage{breakurl}           
\fi

\title{RV32I in ACL2}
\author{Carl Kwan
\institute{The University of Texas at Austin}
\email{carlkwan@cs.utexas.edu}
}
\def\titlerunning{RV32I in ACL2}
\def\authorrunning{Carl Kwan}
\begin{document}
\maketitle

\begin{abstract} 

We present a simple ACL2 simulator for the RISC-V 32-bit base instruction set
architecture, written in the operational semantics style. Like many other ISA
models, our RISC-V state object is a single-threaded object and we prove
read-over-write, write-over-write, writing-the-read, and state well-formedness
theorems. Unlike some other models, we separate the instruction decoding
functions from their semantic counterparts. Accordingly, we verify encoding /
decoding functions for each RV32I instruction, the proofs for which are entirely
automatic. 

\end{abstract}


RISC-V is a popular open-source instruction set architecture (ISA) designed to
be simple, flexible, and scalable. Unlike proprietary ISAs, RISC-V is free to
use and modify, facilitating wide adoption  across industries. A 2022
report suggests ``there are more than 10 billion RISC-V cores in the market, and
tens of thousands of engineers working on RISC-V initiatives
globally''~\cite{rv-news}. This motivates our development of a formal RISC-V
simulator: to analyze and ensure the correctness of RISC-V hardware and software
designs.

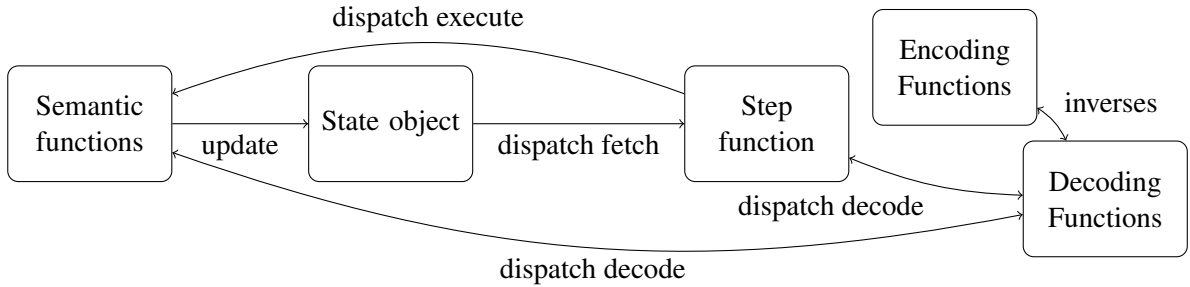
\begin{figure}
\begin{tikzpicture}[scale=1]
 \node[block,draw,fill=white!0] at (0,0) (obj)  {State object};
 \node[block,draw,fill=white!0] at (-4,0) (sem) {Semantic functions};
 \node[block,draw,fill=white!0] at (5,0) (step) {Step function};
 \node[block,draw,fill=white!0] at (9.5,-1) (dis) {Decoding Functions};
 \node[block,draw,fill=white!0] at (7.5,0.75) (asm) {Encoding Functions};
 
\path[->] (obj) edge node[below] {dispatch fetch} (step);
\path[->] (step) edge[bend right=20] node[above] {dispatch execute} (sem);
\path[->] (sem) edge node[below] {update} (obj);
\path[<->] (sem) edge[bend right=15] node[below] {dispatch decode} (dis);
\path[<->] (step) edge[bend right=10] node[below left] {dispatch decode} (dis);
\path[<->] (asm) edge[bend left=15] node[above right] {inverses} (dis);
\end{tikzpicture}
\caption{Overview of the ACL2 RV32I model.\label{fig:overview}}
\end{figure}

We present an executable ACL2 formal model of the 32-bit RISC-V base instruction
set architecture (RV32I)~\cite{risc-v-isa}, formalized by way of operational semantics~\cite{ray2004,ray2008}, and consisting of:
\begin{itemize}
\item a state object, formalized as an ACL2 single-threaded object (stobj)~\cite{stobjs-1,stobj};
\item instruction semantic functions for all 37 RV32I (non-environment) instructions;
\item step / run functions for simulating one or more fetch-decode-execute cycles;
\item standard read-over-write, write-over-write, writing-the-read, and state well-formedness theorems;
\item instruction encoding / decoding functions, and their inversion proofs; 
\item memory conversion theorems for execution using a byte-addressable model
and proving theorems in word-addressable contexts.
\end{itemize}
\Cref{fig:overview} summarizes the executable components in our model.
Our stobj state object \texttt{rv32} consists of:
\begin{itemize}
\item 32 registers, one of which is hardwired to 0 and 31 general-purpose registers;
\item 1 program counter register to hold the address of the current
    instruction;
\item $2^{32}$ bytes of addressable memory; 
\item a model state parameter \texttt{ms} used for debugging
(not an official part of the RISC-V specification).
\end{itemize}
We prove a standard collection of theorems involving the behaviour of
\texttt{rv32} under its access and update functions. These are the read-over-write,
write-over-write, writing-the-read, and state well-formedness
theorems. A more comprehensive treatment of these theorems can be found in the
description of the ACL2 x86 simulator~\cite[p.\ 37]{Goel}, so we discuss
only one example of read-over-write in this document. Reading a byte of memory
in \texttt{rv32} at address \texttt i is made by \texttt{(rm08 i rv32)};
updating \texttt{rv32} at the same memory address with value \texttt{v} is
performed by \texttt{(wm08 i v rv32)}, which returns a new state object. The
following read-over-write theorem
roughly states that if we read a byte from a 32-bit memory address \texttt i
after updating memory address \texttt i with a 8-bit value \texttt v, then we
obtain \texttt v (regardless of what value was at address \texttt i previously):
\begin{lstlisting}[language=acl2]
(defthm rm08-wm08 (implies (and (n32p i) (n08p v)) (equal (rm08 i (wm08 i v rv32)) v)))
\end{lstlisting}
Similar theorems, for every standardized theorem sort, are proven for every
parameter of \texttt{rv32}.

The RISC-V specification intends a byte-addressable memory model; however,
because RV32I instructions (and many other 32-bit extensions) are standardized
to 32-bits and required to be aligned on a four-byte boundary, it is sometimes
easier to reason about memory as if it were word-addressable. Our base model
uses a byte-addressable memory model, but we also formalize and verify functions
for accessing memory as if it were word-addressable. For example, \texttt{rm32}
is a function defined similarly to \texttt{rm08}, but directly obtains 4 bytes
using state access functions. The following theorem states that reading a word
from memory at address \texttt{addr} is equivalent to reading 4 successive bytes
using \texttt{rm08} starting at \texttt{addr} and concatenating the result:
\begin{lstlisting}[language=acl2]
(defthmd rm32-from-successive-bytes
 (equal (rm32 addr rv32)                                   ;; read a word at addr
        (n32 (logior      (rm08      addr   rv32)          ;; read a byte at addr
                     (ash (rm08 (+ 1 addr)  rv32)  8)      ;; read a byte at addr + 1 and shift  8 bits
                     (ash (rm08 (+ 2 addr)  rv32) 16)      ;; read a byte at addr + 2 and shift 16 bits
                     (ash (rm08 (+ 3 addr)  rv32) 24)))))  ;; read a byte at addr + 3 and shift 24 bits
\end{lstlisting}
This enables us to treat the memory in our
\texttt{rv32} state object as if it were
word-addressable, but remain logically consistent to a byte-addressable model.



To simulate the execution of an RV32I instruction, we define instruction
semantic functions, which directly update the \texttt{rv32} state object. These
functions are called by a ``step'' function (see snippet below) that performs the ``fetch'' stage by
obtaining the instruction to be executed from the memory of \texttt{rv32}:
\begin{lstlisting}[language=acl2]
(define rv32-step ((rv32 rv32p))    ;; Takes an rv32 machine state object
 (b* ((PC     (xpc     rv32))       ;; Fetch PC
      (instr  (rm32 PC rv32))       ;; Fetch instruction (32-bit value) from memory
      (opcode (get-opcode instr))   ;; Decode opcode from instruction
      (funct3 (get-funct3 instr))   ;; Decode funct3 from instruction
      (funct7 (get-funct7 instr)))  ;; Decode funct7 from instruction
     (case opcode                   ;; Pattern match on opcode
      (#b0110011                    ;; opcode for R-type instructions
       (case funct3                 ;; Pattern match on funct3
        (#x0                        ;; funct3 for integer ADD / SUB instructions
         (case funct7               ;; Pattern match on funct7
          (#x0  (rv32-add rv32))    ;; funct7 for ADD instruction, offload to rv32-add semantic function
      ...                           ;; Pattern matching and semantic function calls for other instructions
\end{lstlisting}
Note that the step function offloads the decoding of the instruction's
\texttt{opcode}, \texttt{funct3}, and \texttt{funct7} (if applicable), which are
fields that uniquely determine the instruction to be executed, to a layer of
decoding functions (e.g.\ \texttt{get-opcode}, etc.). Actual execution is
dispatched to the instruction semantic functions. Similarly to
\texttt{rv32-step}, semantic functions also offload the decoding of any
registers, memory addresses, or immediate values involved to more decoding
functions. Finally, semantic functions update the \texttt{rv32} state
accordingly.

\begin{figure}
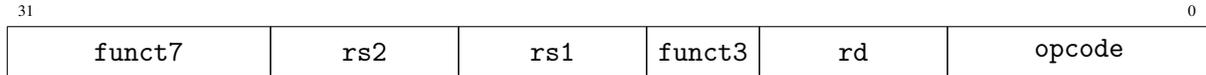

\centering
\begin{bytefield}[bitwidth=\linewidth/32]{32}
\bitheader[endianness=big]{0,31} \\
\bitbox{7}{\texttt{funct7}}
\bitbox{5}{\texttt{rs2}}
\bitbox{5}{\texttt{rs1}}
\bitbox{3}{\texttt{funct3}}
\bitbox{5}{\texttt{rd}}
\bitbox{7}{\texttt{opcode}}
\end{bytefield}
\caption{RV32I R-type instruction format.\label{fig:r-type}}
\end{figure}

There are 6 core instruction formats (R-type, I-type, S-type, B-type, U-type,
and J-type). The format for one of them (R-type) is visualized by
\Cref{fig:r-type}. These formats dictate the role of the particular bits within
an instruction. We formalize decoding functions for obtaining the
appropriate bits as part of the decode stage, e.g.\ the function call
\texttt{(get-opcode instr)} obtains bits 0--7 from the 32-bit value
\texttt{instr}. Other types of instructions may have differing fields and field
sizes, for which we also formalize decoding functions. 
Conversely, we also formalize encoding functions for each RV32I instruction,
e.g.\ \texttt{(asm-add rs1 rs2 rd)} assembles the 32-bit instruction which
stores the sum of the values from registers \texttt{rs1} and \texttt{rs2} into the
destination register \texttt{rd}. A combination of GL~\cite{Swords-gl-1} and simple
rewrite rules enables us to prove ``inverse'' properties, e.g.\ the following
theorem ``recovers'' 
the destination register from an \texttt{asm-add} call:
\begin{lstlisting}[language=acl2]
(defthm get-rd-of-asm-add (equal (get-rd (asm-add rs1 rs2 rd)) (n05 rd)))
\end{lstlisting}
Theorems of this sort are proven for every field (i.e.\ \texttt{funct7},
\texttt{rs2}, \texttt{rs1}, \texttt{funct3}, \texttt{rd}, \texttt{opcode}, and
all \texttt{imm} variations) of every RV32I instruction. Thus, even though calls
are performed using \texttt{get-opcode}, \texttt{get-funct3}, and
\texttt{get-funct7} early within our ``step'' function, proving the correctness
of RV32I instructions does not rely on opening these decoding functions.
Similarly, calls within instruction semantic functions are made to some subset
of \texttt{get-rs1}, \texttt{get-rs2}, \texttt{get-rd}, and various
``get immediate'' functions, but these decoding functions are almost always
disabled. This approach enables us to readily verify the effects of every RV32I instruction on the \texttt{rv32} state. For example, the following theorem determines a priori
the state of an \texttt{rv32} object with a PC pointing to the beginning of an
``add'' instruction after one fetch-decode-execute cycle:
\begin{lstlisting}[language=acl2]
(defthm rv32-step-asm-add-correctness
 (implies (and (rv32p rv32)                                                   ;; rv32 is well-formed
               (< (xpc rv32) *2^32-5*)                                        ;; PC within memory bounds
               (not (equal (n05 k) 0))                                        ;; dest reg is not x0
               (equal (rm32 (xpc rv32) rv32) (asm-add i j k)))                ;; ADD instruction at PC
          (equal (rv32-step rv32)                                             ;; execute 1 CPU cycle 
                 (!xpc (+ (xpc rv32) 4)                                       ;; update PC 
                       (!rgfi (n05 k)                                         
                              (n32+ (rgfi (n05 i) rv32) (rgfi (n05 j) rv32))  ;; reg[k] <- reg[i] + reg[j]
                              rv32)))))                                       
\end{lstlisting}
The upshot is that we now have a verified (with respect to a cycle of
the RISC-V CPU) encoding / semantic function pair for each RISC-V instruction, the
proofs for which are entirely automatic.

Some previous machine models in the operational semantics tradition perform all
the decoding at the top-level within the step and instruction semantic functions
(e.g.\ a single semantic function may decode by bit twiddling without
dispatching to another function) making theorems about a single cycle dependent
entirely on opening a single complicated function. This can hinder verification
efforts for programs whose inputs are abstracted away or not yet known (e.g.\
free variables representing immediate values). A slight novelty in this model is that
we offload all the decoding to a decoding-specific "layer" of functions;
explicitly, we call instruction decoding functions within the step function (see
code snippet for \texttt{rv32-step} above) and the instruction semantic
functions. We prove relevant theorems for these decoding functions so that
future RV32I program verification efforts are more amenable. It is much easier
to prove theorems about pure machine code / bitvectors without the burden of a
CPU structure; the ACL2 code for \texttt{get-rd-of-asm-add} above is an example
of such a theorem. Similarly, it is much easier to prove theorems about a pure
CPU structure without having to worry about bit twiddling; for example, proving
the theorem \texttt{rv32-step-asm-add-correctness} in the previous paragraph
does not involve opening any instruction encoding / decoding functions but instead relies on the
encoding / decoding inversion rules. Connecting the two layers
enables us to prove the desired theorems about the full fetch-decode-execute
cycle by reducing to theorems already proven about the individual layers. 

Our RV32I model is highly inspired by similar ACL2 work for the Y86~\cite{y86},
CHERI-Y86~\cite{cheri-y86}, and x86~\cite{Goel} ISAs. This
work is also partially motivated by the recent development of zero-knowledge virtual
machines (i.e.\ virtual machines which enable one party to prove properties
about a program trace to another party without revealing certain
information, such as the program inputs) for RISC-V programs~\cite{kwan2024}.
One future direction is to develop a verified assembler for RISC-V assembly into
machine code.  Note that our instruction encoding functions, (e.g.\
\texttt{asm-add}) is very near to an assembly function that might parse a string
specifying a RISC-V instruction (e.g.\ integer addition) and return the output
of our encoding function. Furthermore, we are interested in theorems such as
\texttt{rv32-step-asm-add-correctness} because it may be easier to formalize a more
general theorem for an assembler with respect to a CPU cycle when instructions
involve free variables, such as in the function call \texttt{(asm-add i j k)}.
This is in contrast to code proofs, where symbolic execution involving explicit
constants can be common. Another direction for future work is to continue
modelling other 32- or 64-bit RISC-V extensions. On one hand, our experience in
formalizing and verifying the base RV32 instructions involved many repetitive
tasks, suggesting future RISC-V extensions to this model and accompanying
theorems can be easily synthesized, by way of macros or otherwise. On another
hand, our current memory is modelled as a single stobj array, which is
manageable for RV32 but not for RV64. We must improve the resource usage of our
memory model, perhaps by using abstract stobjs similar to how the
bigmem~\cite{bigmem-1,bigmem-2} project is implemented, before tackling the
RISC-V 64-bit instruction set. While there is much future work to be done, we are
optimistic that the application of our model to the formal analysis and
verification of assemblers, programs, hardware, virtual machines, and other
artifacts will result in more reliable RISC-V
infrastructure.

\bibliographystyle{eptcs}
\sloppy
\bibliography{refs}
\end{document}